\documentclass[11pt]{article}

\usepackage{vietnam}
\usepackage{amsmath}
\usepackage{graphicx}
\usepackage{xspace}
\usepackage{bm}

\def\Journal#1#2#3#4{{#1} {\bf #2}, #3 (#4)}

\def\PLB{{\em Phys. Lett.}  B}

\def\PRD{{\em Phys. Rev.} D}

\def\JCAP{\em JCAP} 
\def\JHEP{\em JHEP} 

\def\Contemp{\em Contemp. Phys.}
\def\Jetp{\em JETP Lett.}
\def\LNP{\em Lect. Notes Phys.}

\def\be{\begin{equation}}
\def\ee{\end{equation}}
\def\bea{\begin{eqnarray}}
\def\eea{\end{eqnarray}}

\newcommand{\ie}{i.e.\xspace}

\newlength{\wsingfig}
\newlength{\wdblefig}
\newlength{\wfull}
\newlength{\hfull}
\newcommand{\boldmathsymbol}[1]{{\ensuremath{\boldsymbol{#1}}}}
\newcommand{\bmk}{\boldmathsymbol{k}}

\newcommand{\sss}[1]{{\scriptscriptstyle{#1}}}

\newcommand{\usssE}{\sss{\mathrm{E}}}

\newcommand{\calP}{\mathcal{P}}

\newcommand{\calL}{\mathcal{L}}

\newcommand{\calE}{\mathcal{E}}
\newcommand{\calM}{\mathcal{M}}
\newcommand{\calMref}{\mathcal{M}_{_{\mathrm{REF}}}}
\newcommand{\calMbest}{\mathcal{M}_{_{\mathrm{BEST}}}}

\newcommand{\Rea}{\Re \mathrm{e}\,}

\newcommand{\dd}{\mathrm{d}} 

\newcommand{\Mpc}{\mathrm{Mpc}}

\newcommand{\tetai}{\theta_{\mathrm{inf}}}

\newcommand{\Mp}{M_{_\mathrm{Pl}}}

\newcommand{\Nend}{N_{\mathrm{end}}}
\newcommand{\Nreh}{N_{\mathrm{reh}}}

\newcommand{\EI}{\textit{Encyclop\ae dia Inflationaris}\xspace}

\newcommand{\kstar}{k_*}

\newcommand{\fnlloc}{f_{_{\mathrm{NL}}}^{\mathrm{loc}}}
\newcommand{\fnleq}{f_{_{\mathrm{NL}}}^{\mathrm{eq}}}
\newcommand{\fnlortho}{f_{_{\mathrm{NL}}}^{\mathrm{ortho}}}
\newcommand{\ns}{n_{_\mathrm{S}}}

\newcommand{\wreh}{w_{\mathrm{reh}}}
\newcommand{\wrehbar}{\overline{w}_{\mathrm{reh}}}

\newcommand{\Rrad}{R_{\mathrm{rad}}}

\newcommand{\rhoend}{\rho_{\mathrm{end}}}

\newcommand{\BayesFactor}[2]{B^{#1}_{#2}}
\newcommand{\Bref}[1]{\BayesFactor{#1}{_{\mathrm{REF}}}}
\newcommand{\Bbest}[1]{\BayesFactor{#1}{_{\mathrm{BEST}}}}

\newcommand{\Nparam}[1]{N_\mathrm{param}^{#1}}
\newcommand{\Nuc}[1]{N_{\rm uc}^{#1}}
\newcommand{\Cb}[1]{C_\mathrm{b}^{#1}}

\newcommand{\Photo}{}

\begin{document}
\vspace*{4cm}
\title{INFLATION AFTER PLANCK: AND THE WINNERS ARE ...}

\author{J\'er\^ome Martin}

\address{Insitut d'Astrophysique de Paris, UMR 7095-CNRS, Universit\'e
  Pierre et Marie Curie, \\ 98bis boulevard Arago, 75014 Paris (France)}

\maketitle

\abstracts{We review the constraints that the recently released Cosmic
  Microwave Background (CMB) Planck data put on inflation and we argue
  that single field slow-roll inflationary scenarios (with minimal
  kinetic term) are favored. Then, within this class of models, by
  means of Bayesian inference, we show how one can rank the scenarios
  according to their performances, leading to the identification of
  ``the best models of inflation''.}

\section{Introduction}
\label{sec:intro}

The theory of
inflation~\cite{Starobinsky:1980te,Guth:1980zm,Mukhanov:1981xt,Linde:1981mu,Starobinsky:1982ee}
is currently the leading paradigm to describe the very early
universe. The basic idea is quite simple: the problems of the
pre-inflationary standard cosmological model are avoided if one
postulates that a phase of accelerated expansion took place, at high
energy, before the hot Big Bang era. If gravity is described by
general relativity, then a negative pressure in the effective stress
energy tensor sourcing the Einstein equations is all we need to
produce this acceleration. Moreover, since, in the situation described
above, field theory is the correct framework to describe matter and
since a preferred direction (\ie a spin or a vector) is not compatible
with homogeneity and isotropy, a scalar field, the so-called inflaton
field, appears to be the ideal candidate. Indeed, in that case, the
pressure is given by the difference between the kinetic and the
potential energy. Therefore, if the potential energy dominates over
the kinetic energy, that is to say if the field slowly rolls down its
potential, then one automatically produces a phase of inflation.

Inflation also naturally leads to a convincing mechanism for structure
formation~\cite{Mukhanov:1981xt,Starobinsky:1982ee} and this is
probably the reason why this scenario is considered as very
attractive. In brief, the quantum fluctuations of the coupled inflaton
and gravitational fields are amplified and give rise to an almost
scale invariant power spectrum in full agreement with the
astrophysical observations. An attractive feature of this mechanism is
that it is quite conservative: it is nothing but particle creation
(the quantized cosmological perturbations) under the influence of a
classical source (the background gravitational field). This is
well-known in quantum field theory and is in fact the essence of the
so-called Schwinger effect~\cite{Martin:2007bw}.

Although pretty straightforward regarding the physical principles,
inflation turns out to be more complicated when it comes to concrete
implementation. Indeed, there are literally hundreds of different
models of inflation depending on whether there is one or several
fields, with minimal or non-minimal kinetic terms, and/or with a
featureless or not potential. In addition, all the possible
combinations (for instance, several scalar fields with non-minimal
kinetic terms) are also possible. How, then, can we identify to which
version of inflation we are dealing with?

A priori, one could solely use theoretical considerations based on
high energy physics to single out a unique consistent model. This
seems to be unrealistic today since, at the energy scales relevant for
inflation, particle physics is not known and remains
speculative. Moreover, the fact that we deal with so many models
precisely originates from the fact that many possible versions of BSM
(Beyond the Standard Model) physics exist leading to a plethora of
different inflationary scenarios. For instance, models with a Dirac
Born Infeld (DBI) kinetic term have been considered because this
specific case can be motivated by string theory.

This leaves us with a ``mixed approach'' which consists, from the
currently available scenarios and from the data, in inferring the
correct model of inflation. In these proceedings, we explore this
route and discuss the consequences for inflation of the recently
released Planck data~\cite{Planck,PlanckNG}. These data tell us that
we live in a spatially flat universe, $100 \Omega
_K=-0.05{}^{+0.65}_{-0.66}$, which is of course very consistent with
inflation and that the cosmological fluctuations are adiabatic (at
$95\%$ CL) and Gaussian $\fnlloc=2.7\pm 5.8$, $\fnleq=-42\pm 75$ and
$\fnlortho=-25\pm 39$~\cite{PlanckNG}. Another important message of
the Planck data~\cite{Planck} is the fact that a tilt in the power
spectrum has now been detected at a significant statistical level,
$\ns=0.9603\pm 0.0073$, thus ruling out scale invariance at more than
$5\sigma$. In addition, neither a significant running nor a
significant running of the running have been detected since it was
found that $\dd \ns /\dd \ln k=-0.0134\pm 0.009$ (Planck+WP) and $\dd
^2 \ns/\dd \ln^2 k=0.02\pm 0.016 $ (WMAP+WP), with a pivot scale
chosen at $\kstar=0.05 \Mpc^{-1}$.

Based on the above discussion, it is clear that single field slow roll
models (with a minimal kinetic term) are favored from an
observational point of view since this class of models precisely
predicts no entropy perturbations and negligible non-Gaussianities. Of
course, this does not mean that other inflationary scenarios are ruled
out but simply that there are not needed to explain the
data. Inflation therefore appears as a simple and non trivial, but non
exotic, theory.

It should however be clear that, even if we restrict our
considerations to this simple class of models, it still remains a very
large number of possible models~\cite{Martin:2013tda}. Then comes the
questions of how one can constrain these models, estimate their
performances and rank them, in a statistically well-defined fashion in
order to find ``the best model(s) of inflation''. Once a well
justified method has been designed, it can be applied to all
inflationary models in order to actually identify which scenario is
favored by the Planck data. Answering and discussing these questions
is the main subject of the present paper.

This article is organized as follows. In the next section,
Sec.~\ref{sec:slowroll}, we briefly review slow-roll inflation. Then,
in Sec.~\ref{sec:ranking}, we define and discuss what is meant by a
model A is better than a model B. For this purpose, we review the
Bayesian model comparison approach, we quickly recall how the Bayesian
evidence of a slow-roll inflationary model can be estimated and we
present the results of Ref.~\cite{Martin2013} which give the model
winners. Finally, in the conclusion, Sec.~\ref{sec:conclusions}, we
summarize our results.

\section{Slow-Roll Inflation and CMB Measurements}
\label{sec:slowroll}

Slow-roll inflation is a very simple system. It consists in one scalar
field with a minimal kinetic term and a potential $V(\phi)$ and its
behavior is controlled by the Friedmann-Lema\^{\i}tre and
Klein-Gordon equations, namely
\begin{align}
\label{eq:kgfriedman}
H^2 &=\frac{1}{3\Mp^2}\left[\frac{\dot{\phi}^2}{2}+V(\phi)\right], \qquad
\ddot{\phi}+3H\dot{\phi}+V_{\phi} = 0,
\end{align}
where $H\equiv \dot{a}/a$ denotes the Hubble parameter, $a(t)$ being
the Friedmann-Lema\^{\i}tre-Robertson Walker (FLRW) scale factor and
$\dot{a}$ its derivative with respect to cosmic time $t$. $\Mp = 8 \pi
G$ denotes the reduced Planck mass. A subscript $\phi$ means a
derivative with respect to the inflaton field. Therefore, the only
unknown function is the potential and, here, we try to constrain its
shape using the Planck data.

When the potential is no longer flat enough (this usually happens when
the system approaches its ground state, \ie the minimum of the
potential), inflation stops, the inflaton field
decays~\cite{Turner:1983he,Kofman:1997yn}, the decay products
thermalize~\cite{Podolsky:2005bw} and this is how inflation is
smoothly connected to the standard hot Big Bang phase. Let $\rho $ and
$P$ be the energy density and pressure of the effective fluid
dominating the Universe during reheating and $\wreh\equiv P/\rho$ the
corresponding ``instantaneous'' equation of state. One can also define
the mean equation of state parameter, $\wrehbar$,
by~\cite{Martin:2010kz}
\begin{equation} 
\wrehbar \equiv \frac{1}{\Delta N}\int_{\Nend}^{\Nreh} \wreh(n)\dd n,
\end{equation}
where $\Delta N \equiv \Nreh - \Nend$ is the total number of e-folds
during reheating, $\Nend$ being the number of e-folds at the end of
inflation and $\Nreh$ being the number of e-folds at which reheating
is completed and the radiation dominated era begins. Then, one
introduces a new parameter~\cite{Martin:2010kz}
\begin{equation}
\label{eq:RradN} 
\ln \Rrad \equiv \frac{\Delta N}{4}\left(-1+3\, \wrehbar\right).
\end{equation} 
As discussed in detail in Ref.~\cite{Martin:2010kz}, this parameter
completely characterizes the reheating phase and its knowledge is
necessary in order to work out the inflationary predictions for the
CMB. In particular, it can be related to the so-called reheating
temperature through~\cite{Martin:2010kz}
\begin{equation}
T_{\mathrm{reh}}^4=\frac{30\rhoend}{\pi^2g_\star}\Rrad^{12(1+\wrehbar)/(1-3\wrehbar)},
\end{equation}
where $\rhoend$ is the energy density at the end of inflation, which
is known when $V(\phi)$ has been chosen, and $g_\star$ is the number
of degrees of freedom at that time.

Let us now turn to the description of inflationary perturbations. Two
types of fluctuations are relevant for inflation: density
perturbations and primordial gravity waves. The density perturbations
are described in terms of the Mukhanov-Sasaki variable $v(\eta,{\bm
  x})$. In the Schr\"odinger approach, the quantum state of the system
is described by a wavefunctional, $\Psi\left[v(\eta,{\bm x})\right]$,
which can be factorized into mode components as~\cite{Martin:2012pea}
\begin{equation}
\Psi\left[v(\eta,{\bm x})\right]=\prod _{\bm k}
\Psi_{\bm k}\left(v_{\bm{k}}^\mathrm{R},
v_{\bm{k}}^\mathrm{I}\right)
=\prod_{\bm k}\Psi^{\rm R}_{\bm k}\left(v_{\bm{k}}^\mathrm{R}\right)
\Psi ^{\rm I}_{\bm k}\left(v_{\bm{k}}^\mathrm{I}\right),
\end{equation}
where $v_{\bm{k}}^\mathrm{R}$ denotes the real part of $v$ and
$v_{\bm{k}}^\mathrm{I}$ its imaginary part. Each wavefunction obeys a
Schr\"odinger equation with an Hamiltonian that can be deduced from a
second order expansion of the action ``gravity + inflaton
field''. Then, one can show that the solution is explicitly
time-dependent and given by a Gaussian ($\eta $ being the conformal
time)
\begin{equation}
\label{eq:gaussianpsi}
\Psi_{\bm k}^{\rm R,I}\left(\eta,v_{\bm k}^{\rm R,I}\right)
=N_{\bm k}(\eta){\rm e}^{-\Omega_{\bm k}(\eta)\left(v_{\bm k}^{\rm R,I}\right)^2}.
\end{equation}
where the functions $N_{\bm k}(\eta)$ and $\Omega _{\bm k}(\eta)$ can
be expressed as~\cite{Martin:2012pea}
\begin{equation}
\label{eq:solpsi}
\left \vert N_{\bm k}\right \vert 
=\left(\frac{2\Rea  \Omega _{\bm k}}{\pi}\right)^{1/4}, \quad 
\Omega_{\bm k}=-\frac{i}{2}\frac{f_{\bm k}'}{f_{\bm k}}.
\end{equation}
The function $f_{\bm k}$ obeys the equation of motion of a parametric
oscillator, namely $f_{\bm k}''+\omega^2f_{\bm k}=0$, where the time
dependent frequency of this oscillator is given by
$\omega^2\left(\eta, \bm{k}\right)=k^2-
\left(a\sqrt{\epsilon_1}\right)^{\prime\prime}/(a\sqrt{\epsilon_1})$,
$k$ being the wavenumber of the mode under consideration and
$\epsilon_1\equiv -\dot{H}/H^2$ the first slow-roll parameter
characterizing the cosmological expansion during inflation. For
gravitational waves, one also obtains a Gaussian wave-function except
that the fundamental frequency of the oscillator $f_{\bm k}$ is now
given by $\omega^2=k^2-a''/a$.

One of the great advantage of inflation is that it is possible to
choose well justified initial conditions. In brief, this is because,
at the beginning of inflation, the physical wavelengths of Fourier
modes of cosmological relevance today are much smaller than the Hubble
radius. These modes do not feel spacetime expansion and, as a
consequence, it is natural to choose the vacuum state as their initial
state. Technically, this amounts to take $\Omega_{\bm k}=k/2$ at
initial time in Eq.~(\ref{eq:solpsi}) which indeed corresponds to the
ground state wavefunction of an harmonic oscillator.

We have just seen that the effective frequency of density
perturbations depends on the first slow-roll parameter and its
derivatives. For this reason, it is interesting to define a hierarchy
of slow-roll parameters by means of the following
formula~\cite{Schwarz:2004tz}
\begin{equation}
\label{eq:defhf}
\epsilon_{n+1} \equiv \frac{\dd \ln \left \vert 
\epsilon_n \right \vert}{\dd N}, 
\quad n\ge 0,
\end{equation}
where $\epsilon_0\equiv H_{\mathrm{ini}}/H$. The slow-roll conditions
refer to a situation where all the $\epsilon_n$'s satisfy
$\epsilon_n\ll 1$. From this definition, we see that $\omega(k,\eta)$
for density perturbations depends on $\epsilon_1$, $\epsilon_2$ and
$\epsilon_3$ while, for gravity waves, it only depends on
$\epsilon_1$. Notice that, since $H(\phi)$ and $V(\phi)$ are related
through the Einstein equations, the parameters $\epsilon_n$ can also
be expressed in terms of the successive derivatives of the potential,
namely
\begin{align} 
\label{eq:eps1}
\epsilon_1 & \simeq
\frac{\Mp^2}{2}\left(\frac{V_\phi}{V}\right)^2, \\ 
\label{eq:eps2}
\epsilon_2 & \simeq
2\Mp^2\left[\left(\frac{V_\phi}{V}\right)^2-\frac{V_{\phi \phi}}{V}\right], \\
\label{eq:eps3}
\epsilon_2\epsilon_3 & \simeq 2\Mp^4\left[
\frac{V_{\phi \phi \phi}V_\phi}{V^2}-3\frac{V_{\phi \phi}}{V}
\left(\frac{V_\phi}{V}\right)^2
+2\left(\frac{V_\phi}{V}\right)^4\right].
\end{align} 

The slow-roll approximation also allows us to solve the equation that
controls the evolution of the function $f_{\bm k}$ and, therefore, of
the wavefunction. Since the initial conditions are also completely
specified (see the above discussion), the function $f_{\bm k}$ and,
hence, the wavefunction, is completely known. One can then calculate
the two-point correlation function of the Mukhanov-Sasaki variable or,
in Fourier space, of the power spectrum\footnote{For density
  perturbations, the definition of the power spectrum reads
\begin{equation} 
\label{Pzeta}
\calP_{\zeta}(k)\equiv \frac{k^3}{4\pi^2 \Mp^2}\left\vert
  \frac{v_\bmk}{a\sqrt{\epsilon _1}}\right\vert ^2\, .
\end{equation}}. This involves a double expansion. The
power spectrum is first expanded around a chosen pivot scale $\kstar$ such that
\begin{equation} 
\label{spectrumsr}
\frac{\calP(k)}{\calP_0} = a_0 + a_1 \ln \left(\dfrac{k}{\kstar}\right) 
 + \frac{a_2}{2} \ln^2\left(\dfrac{k}{\kstar}\right)
 + \dots \, ,
\end{equation}
where $ \calP_{\zeta 0} =H^2/\left(8 \pi^2 \epsilon_1 \Mp^2\right)$
and, then, the coefficients $a_n$ are expanded in terms of the
slow-roll parameters. Concretely, for scalar perturbations, at second
order in the slow-roll approximation, one
obtains~\cite{Schwarz:2004tz,Martin:2013uma}
\begin{eqnarray}
\label{eqn:as0}
a_{0} &=& 1 - 2\left(C + 1\right)\epsilon_{1*} - C \epsilon_{2*}
+ \left(2C^2 + 2C + \frac{\pi^2}{2} - 5\right) \epsilon_{1*}^2 \nonumber
\\ & + & \left(C^2 - C + \frac{7\pi^2}{12} - 7\right)
\epsilon_{1*}\epsilon_{2*} + \left(\frac12 C^2 + \frac{\pi^2}{8} -
1\right)\epsilon_{2*}^2 \nonumber \\ & + & \left(-\frac12 C^2  +
\frac{\pi^2}{24}\right)  \epsilon_{2*}\epsilon_{3*} \, , \\
\label{eqn:as1}
a_{1} & = & - 2\epsilon_{1*} - \epsilon_{2*} + 2(2C+1)\epsilon_{1*}^2
+ (2C - 1)\epsilon_{1*}\epsilon_{2*} + C\epsilon_{2*}^2 - C\epsilon_{2*}\epsilon_{3*}
\, ,\\ 
a_{2}  &=& 4\epsilon_{1*}^2 + 2\epsilon_{1*}\epsilon_{2*} +
\epsilon_{2*}^2 - \epsilon_{2*}\epsilon_{3*} \, ,
\label{eqn:as2}
\end{eqnarray}
where $C \equiv \gamma_{\usssE} + \ln 2 - 2 \approx -0.7296$, $\gamma
_\usssE$ being the Euler constant. $\epsilon_{n*}$ denotes the value
of the function $\epsilon_n$ at Hubble radius crossing during
inflation. For gravitational waves, the power spectrum has the same
structure but the expressions of the coefficients $a_n$ differ.

In order to make concrete predictions, we must calculate the numerical
values of the quantities $\epsilon_{n*}$. In order to do so, one needs
to know the slow-roll trajectory and we need to calculate accurately
when inflation stops. As a result, $\epsilon_{n*}$ usually depends on
$\tetai$, the parameters of the potential $V(\phi)$, and on the
reheating temperature:
$\epsilon_{n*}=\epsilon_{n*}(\tetai,T_{\mathrm{reh}})$.

The above considerations explain how the CMB can tell us something
about inflation. Indeed, CMB measurements constrain the power
spectrum, that is the say, given the form the expression of $\calP(k)$
above, the values of the parameters
$\epsilon_{n*}(\tetai,T_{\mathrm{reh}})$. These parameters carry
information about the shape of the potential (recall the expression of
the slow-roll parameters in terms of the derivative of the potential)
and on the reheating temperature. As a consequence, one can infer what
are the properties of the inflaton potential $V(\phi)$ and learn about
the physical conditions that prevailed in the early universe.

\section{Ranking the Inflationary Models}
\label{sec:ranking}

\subsection{Bayesian Analysis in Brief}
\label{subsec:bayesanalysis}

In the previous section, we have described how one can calculate the
predictions of a given inflationary model. However, we also would like
to compare the performances of the different inflationary scenarios
and one way to achieve this program is to compare the quality of the
fits provided by the different models.

Let us now briefly describe how this can be
achieved~\cite{Kunz:2006mc,Trotta:2008qt,Martin:2010hh}. Let us call
$\calM_1$ and $\calM_2$ two competing models, aiming at explaining
some data $D$ (here, of course, we have in mind the Cosmic Microwave
Background - CMB - measurements), the model one depending on one
parameter, $\theta$, and the model two depending on two parameters,
$\alpha $ and $\beta$. Their likelihood function can be written as
\begin{equation}
\calL_1\left(D\vert
  \theta \right)=\calL_{1,\mathrm{max}}
\mathrm{e}^{-\chi^2\left(\theta\right)/2}, 
\quad\calL_2\left(D\vert
  \alpha,\beta\right)=\calL_{2,\mathrm{max}}
\mathrm{e}^{-\chi^2\left(\alpha,\beta\right)/2},
\end{equation}
where $\chi^2$ is the effective chi-squared of the corresponding model
that we do not need to specify at this stage. The quality of the fits
can be estimated by computing the ratio of the maximums of the two
likelihoods. However, this does not give us information regarding the
complexity of the two models\footnote{In the following, we will
  introduce a quantity called the ``Bayesian complexity''. Here, we
  use the word ``complexity'' in the standard sense, \ie a model is
  more complicated than another if, for instance, it has more
  parameters or more fine-tuning. At this stage, it should not be
  confused with the Bayesian complexity.}. If, for instance, model
$\calM_2$ achieves a very good fit only at the price of a fine-tuning,
while $\calM_1$ ``naturally'' performs well, one may wish to penalize
$\calM_2$ for its complexity. This ``Occam's razor'' criterion is
automatically included if one characterizes a model by its Bayesian
evidence~\cite{Trotta:2008qt}. The Bayesian evidence is the integral
of the likelihood function over the prior space. Concretely, for
$\calM_1$ and $\calM_2$, this leads to
\begin{equation}
\calE_1=\int \calL_1\left(D\vert \theta\right) 
\pi\left(\theta\right)\dd \theta, \quad 
\calE_2=\int \calL_2\left(D\vert \alpha,\beta\right) 
\pi\left(\alpha,\beta\right)
\dd \alpha\dd \beta.
\end{equation}
The prior distributions $\pi(\theta)$ and $\pi(\alpha,\beta)$,
satisfying $\int \pi(\theta)\dd \theta =1$ [and a similar expression
for $\pi\left(\alpha,\beta\right)$], encodes what we know about the
parameter $\theta$ before our information is updated when we learn
about the data $D$. Let us notice that the likelihood functions are
not normalized in the sense that $\int \calL_1\left(D\vert
  \theta\right)\dd \theta\neq 1$. For simplicity, let us now assume
that the prior $\pi(\theta)$ is flat in the range
$[\theta_{\mathrm{min}},\theta_{\mathrm{max}}]$ and vanishes
elsewhere. Because the distribution is normalized, one has
$\pi(\theta)=1/\Delta \theta$ with $\Delta \theta
=\theta_{\mathrm{max}}-\theta_{\mathrm{min}}$. Let us also assume that
the likelihood function has a bell shape (for instance, but
necessarily, is a Gaussian function) characterized by the width
$\delta \theta$. Let us finally suppose that the data give more
information than the prior, in other words that the likelihood is more
peaked than the prior. In that case, the Bayesian evidence of model
$\calM_1$ can be approximated by
\begin{equation}
\label{eq:evidenceone}
\calE_1\simeq \calL_{1,\mathrm{max}}\frac{\delta \theta}{\Delta \theta}.
\end{equation}
In the same fashion, with the same assumptions (and obvious
notations), the evidence of model $\calM_2$ can be expressed as
\begin{equation}
\calE_2\simeq \calL_{2,\mathrm{max}}\frac{\delta \alpha}{\Delta \alpha}
\frac{\delta \beta}{\Delta \beta}.
\end{equation}
Then, applying Bayes' theorem, the probability of model $\calM_1$ is
given by $p(\calM_1\vert D)=\calE_1\pi(\calM_1)/p(D)$ and a similar
formula for $p(\calM_2\vert D)$. In this expression, $\pi(\calM_1)$
represents the prior of model $\calM_1$ and the quantity $p(D)$ is a
normalization factor. If we say that, initially, the two models are
equally probable, that is to say $\pi(\calM_1)=\pi(\calM_2)$, then the
ratio of their posterior probabilities, the so-called Bayes factor,
can be expressed as
\begin{equation}
B_{21}\equiv \frac{p(\calM_2\vert D)}{p(\calM_1\vert D)}=\frac{\calE_2}{\calE_1}
=\frac{\calL_{2,\mathrm{max}}}{\calL_{1,\mathrm{max}}}\frac{\delta \alpha}{\Delta \alpha}
\frac{\delta \beta}{\Delta \beta}\frac{\Delta \theta}{\delta \theta}.
\end{equation}
We see that the Bayes factor is controlled by the ratio
$\calL_{2,\mathrm{max}}/\calL_{1,\mathrm{max}}$ but now weighted by a
factor, the so-called Occam factor, which penalizes the more
complicated model, $\calM_2$, for any wasted parameter space. If, for
instance, we take $\delta \alpha/\Delta \alpha=\delta \beta/\Delta
\beta=\delta \theta/\Delta \theta =0.01$, then
$B_{21}=0.01\calL_{2,\mathrm{max}}/\calL_{1,\mathrm{max}}$ and the
more complicated model can win only if its likelihood at the ``best
fit point'' is two orders of magnitude larger than that of
$\calM_1$. So the best model is the model which can achieve the best
compromise between simplicity and quality of the fit.

From the previous considerations, we see that the Bayesian evidence is
an ideal tool to rank models and to find the best model. Nevertheless,
it has the following property that could be considered as a
shortcomings. Suppose we define a model $\calM_3$ such that it is in
fact model $\calM_2$ but with a third parameter, say $\gamma$, such
that this new parameter does not affect in any way the fit to the
data; in other words, such that the likelihood is flat along $\gamma
$. In that case, the evidence of model $\calM_3$ is given by
\begin{eqnarray}
\calE_3 &=& \int  
\calL_3(D\vert \alpha, \beta,\gamma )\pi(\alpha)\pi(\beta )\pi(\gamma)
\dd\alpha \dd\beta \dd \gamma =\int  
\calL_3(D\vert \alpha, \beta )\pi(\alpha)\pi(\beta )\pi(\gamma)
\dd\alpha \dd\beta \dd \gamma 
\\
&=& \int \calL_2(D\vert \alpha, \beta )\pi(\alpha)\pi(\beta )
\dd\alpha \dd\beta \int \pi(\gamma )\dd \gamma=\calE_2. 
\end{eqnarray}
Therefore, the two models have the same evidence despite the fact that
$\calM_2$ is obviously simpler than $\calM_3$. In order to break this
degeneracy, one has to introduce another quantity, the Bayesian
complexity~\cite{Kunz:2006mc}, which allows us to distinguish
$\calM_2$ and $\calM_3$.

In order to discuss the definition of the complexity, we work with a
one parameter model only, \ie $\calM_1$, (the generalization to an
arbitrary number of parameters is straightforward) and we explicitly
assume that the likelihood of the model is a Gaussian, namely
\begin{equation}
\calL_1\left(D\vert \theta \right) =\calL_{1,\mathrm{max}}
\, \mathrm{e}^{-\left(\theta-d\right)^2/\left(2\sigma ^2\right)},
\end{equation}
where $d$ represents a measurement of the parameter
$\theta$. Regarding the prior, instead of considering a flat
distribution as before, we also assume it is given by a Gaussian
centered at $\theta=\mu$,
\begin{equation}
\pi \left(\theta \right) =\frac{1}{\Sigma \sqrt{2\pi}}
\, \mathrm{e}^{-\left(\theta-\mu \right)^2/\left(2\Sigma ^2\right)}.
\end{equation}
We can check that this distribution is properly normalized. These new
assumptions are made for convenience only and do not change the above
discussion (in fact, not quite exactly, see below). In particular,
now, $\delta \theta$ is clearly given by $\sigma $ and the $\Delta
\theta $ by $\Sigma$ so that the condition that the data are more
informative than the prior, $\delta \theta \ll \Delta \theta$,
corresponds to $\sigma \ll \Sigma $. Then one can calculate the
posterior distribution of the parameter $\theta$,
\begin{eqnarray}
\label{eq:postdistri}
p\left(\theta \vert D\right) &=&
\frac{1}{\calE_1}\calL_1\left(D\vert \theta\right)
\pi\left(\theta\right)\\
&=&\frac{1}{\sqrt{2\pi}}\sqrt{\frac{1}{\Sigma ^2}+\frac{1}{\sigma^2}}
\exp\left[-\frac{1}{2}\left(\frac{1}{\Sigma ^2}+\frac{1}{\sigma ^2}
\right)\left(\theta-\frac{d+\mu_1\sigma ^2/\Sigma ^2}{1+\sigma ^2/\Sigma ^2}
\right)^2\right],
\end{eqnarray}
which is a properly normalized Gaussian with mean and variance
respectively given by
\begin{equation}
\frac{d+\mu \sigma ^2/\Sigma ^2}{1+\sigma ^2/\Sigma ^2}, \qquad 
\frac{1}{\sqrt{\frac{1}{\Sigma ^2}+\frac{1}{\sigma ^2}}}.
\end{equation}
On the other hand, the evidence of the model can be expressed as
\begin{equation}
\calE_1=\frac{\calL_{1,\mathrm{max}}}{\sqrt{1+\Sigma ^2/\sigma ^2}}
\mathrm{e}^{-\left(\mu -d\right)^2/\left[2\left(\sigma^2+\Sigma ^2\right)\right]}.
\end{equation}
This result is compatible with the previous discussion. Indeed, if the
likelihood is more informative than the prior, then $\Sigma/\sigma \gg
1$ and the factor in front of the exponential reads $\sim
\calL_{1,\mathrm{max}}\sigma/\Sigma$ which is equivalent to
$\calL_{1,\mathrm{max}}\delta \theta/\Delta \theta$ and shows that the
Occam's factor is simply $\sigma/\Sigma$.

We now come to the definition of the Bayesian complexity denoted by
$C_\mathrm{b}$ in what follows. It reads~\cite{Kunz:2006mc}
\begin{equation}
C_\mathrm{b}=\left\langle \chi^2\left(\theta \right)\right \rangle 
-\chi^2\left(\left \langle \theta \right \rangle \right),
\end{equation}
where the symbol $\langle \cdots \rangle $ means an average of the
quantity $\cdots$ with a weigh given by the posterior $p(\theta\vert
D)$. In the above expression, the effective $\chi^2$ is defined by
$-2\ln \calL$, which in the present case, reads
\begin{equation}
  \chi^2(\theta)=\frac{1}{\sigma^2}\left(\theta-d\right)^2
-2 \ln \calL_{1,\mathrm{max}}.
\end{equation}
Then, using the explicit expression for the posterior distribution,
see Eq.~(\ref{eq:postdistri}), and the previous expression for the
$\chi^2$, one obtains the following formula for the Bayesian
complexity
\begin{equation}
C_\mathrm{b}
=\int p(\theta\vert D)\chi^2(\theta)\dd \theta 
-\chi^2\left[\int p(\theta\vert D)\theta \dd \theta\right]
=\frac{1}{1+\sigma^2/\Sigma^2}.
\end{equation}
Therefore, if $\sigma \ll \Sigma$, one has $C_\mathrm{b}\simeq 1$. In
other words, since the likelihood function is much more peaked than
the prior, the parameter $\theta $ is well-measured and the complexity
is one. If, one the contrary, $\sigma \gg \Sigma$, then
$C_\mathrm{b}\simeq 0$ and the data are not accurate enough to
constraint $\theta$. In the multidimensional case (\ie a model with $n$
parameters), one has $C_\mathrm{b} =
\sum_{i=1}^n1/(1+\sigma_i^2/\Sigma_i^2)$, and the complexity gives the
number of parameters that have been measured with the data $D$ or, in
other words, the number of eigendirections in which the likelihood is
more informative than the prior.

Finally, to conclude this section, let us try to derive the complexity
for another very simple one parameter model, similar to the example we
treated at the beginning of this article. This will help us to
understand the meaning of complexity in another
context~\cite{Feroz:2011bj}. We assume that the likelihood is flat,
centered at $\theta=0$ with a width given by $\delta \theta$ and a
height $\calL_\mathrm{max}$. We also assume that the prior is flat in
the range $[-\Delta \theta/2, \Delta \theta/2]$ and has height
$1/\Delta \theta$ (and is less informative than the likelihood). In
that case, it is straightforward to estimate the evidence of the model
which is $\calE=\calL_\mathrm{max}\delta\theta /\Delta \theta$. On the
other hand, the posterior on the parameter $\theta$ can be expressed
as
\begin{equation}
p\left(\theta\vert D\right)=\frac{\calL_{\mathrm{max}}}{\Delta \theta \calE}
=\frac{1}{\delta \theta}, \quad \mbox{for} 
-\frac{\delta \theta}{2}<\theta <\frac{\delta \theta}{2},
\end{equation}
and vanishes otherwise. As a consequence, one finds that the
complexity can be written as
\begin{equation}
C_\mathrm{b} =-2\int _{-\delta \theta/2}^{\delta \theta/2} \frac{1}{\delta \theta}
\ln \calL \, \dd \theta +2\ln \calL_{\mathrm{max}}=0.
\end{equation}
We see that one can no longer interpret the complexity as we did
before. The reason is that the model we have used is too far from a
Gaussian model and the concept of complexity cannot be really defined
in that case. This illustrates the limitation of this statistical tool
which is efficient only if the underlying statistics is not too far
from a Gaussian. This is a warning that should be kept in mind in the
following.

\subsection{Inflationary Bayesian Inference}
\label{subsec:infbayes}

\begin{table}[t]
\begin{center}
\begin{tabular}{l l  l} \hline 
  $|\ln \Bref{i}|$ & Odds  & Strength of evidence \\\hline 
 $<1.0$ & $< 3:1$ &  Inconclusive \\
 $1.0$ & $\sim 3:1$ &  Weak evidence \\
 $2.5$ & $\sim 12:1$ & Moderate evidence \\
 $5.0$ & $\sim 150:1$ &  Strong evidence \\
\hline
\end{tabular}
\caption{Jeffreys scale for evaluating the strength of evidence when
  comparing two models, $\calM_i$ versus a reference model
  $\calMref$.}
\label{tab:Jeff} 
\end{center}
\end{table}

Following the above considerations, it should now be clear that one
way to estimate the performances of inflationary models (in explaining
the recently released Planck data) is to calculate their evidence and
their complexity. Then, one can rank them in a statistically
consistent way and find the best scenarios. The predictions of all
single field scenarios have been worked out and compared to Planck
data in \EI~\cite{Martin:2013tda} and the calculation of the evidences
and complexity for those models was performed in
Ref.~\cite{Martin2013} using a method recently developed in
Ref.~\cite{Ringeval2013}. From these results, one can determine the
Bayes factor defined by
\begin{eqnarray}
\Bref{i}  & \equiv & \frac{\calE(D|\calM_i)}{\calE(D|\calMref)}\, ,
\end{eqnarray}
where the reference model was taken to be the Starobinsky model. The
``Jeffreys scale'', see Table~\ref{tab:Jeff}, gives an empirical
prescription for translating the values of $\Bref{i}$ into strengths
of belief. One can summarize our results as follows. Firstly, for
convenience, one can change the reference point of the Bayes factor
and estimate the quantity $\Bbest{i}\equiv
\calE(D|\calM_i)/\calE(D|\calMbest)$ (rather than $\Bref{i}$ before)
with non-committal model priors. Then, one uses the Jeffreys scale
with $\Bbest{i}$, instead of $\Bref{i}$, and count the number of
models in the ``inconclusive'', ``weak evidence'', ``moderate
evidence'' and ``strong evidence'' zones. The models in the
``inconclusive'' category can be viewed as the best models. We have
found that this is the case for $52$ models for a total of $193$
models, that is to say $26\%$ of the models. Therefore, this means
that $\simeq 73\%$ of the inflationary scenarios can now be considered
as disfavored and/or ruled out by the Planck data.

Secondly, one determines the number of unconstrained parameters,
$\Nuc{i}$, which is the number of parameters of model $\calM_i$,
$\Nparam{i}$, minus its complexity $\Cb{i}$
\begin{equation}
\Nuc{i}=\Nparam{i}-\Cb{i}.
\end{equation}
Then, among the models in the ``inconclusive'' region, one should
prefers models for which $\Nuc{i}\simeq 0$. If one retains the
criterion $0<\Nuc{i}<1$, then one reduces the number of ``good
models'' to $17$, that is to say to $\simeq 9\%$ of the \EI scenarios.

\begin{figure}[t]
\begin{center}
\includegraphics[width=12cm]{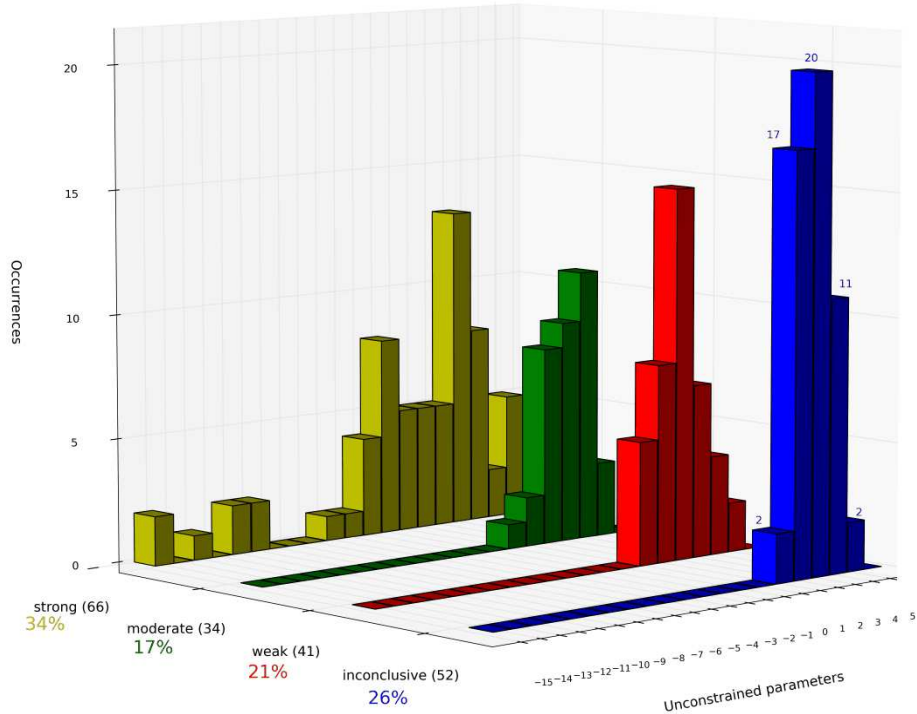}
\caption{Histogram representing the number of inflationary models
  after Planck2013 according to the Jeffrey category and the number of
  unconstrained parameters.}
\label{fig:histo}
\end{center}
\end{figure}

These results are summarized in Fig.~\ref{fig:histo} which shows the
histogram corresponding to the number of models in each Jeffreys
category with a given value of $\Nuc{i}$. A complete analysis and the
list of the best models can be found in Ref.~\cite{Martin2013}.

\section{Conclusions}
\label{sec:conclusions}

In these proceedings, we have analyzed the implications of the
recently released Planck data for inflation. We have argued that
single field slow-roll scenarios with minimal kinetic term are favored
by Planck 2013. Then, we have designed specific Bayesian tools to
further constrain the models within the class of favored scenarios. We
have shown that Planck2013 can then single out about $\sim 10\%$ of
the models, thus strongly reducing the inflationary landscape
compatible with the astrophysical observations. Our results
demonstrate concretely that CMB data can constrain the physics of the
early universe in an efficient way. In the near future, the next
release of Planck measurements should allow us to learn even more
about inflation.

\section*{Acknowledgments}

I would like to thank C.~Ringeval and V.~Vennin for careful reading of
the manuscript.


\begin{thebibliography}{}

\end{thebibliography}


\section*{References}

\begin{thebibliography}{99}

\bibitem{Starobinsky:1980te} A. Starobinsky, \Journal{\PLB}{91}{99}{1980}.

\bibitem{Guth:1980zm} A. Guth, \Journal{\PRD}{23}{347}{1981}.

\bibitem{Mukhanov:1981xt} V. Mukhanov and G. Chibisov, \Journal{\Jetp}{33}{532}{1981}.

\bibitem{Linde:1981mu} A. Linde, \Journal{\PLB}{108}{389}{1982}.

\bibitem{Starobinsky:1982ee} A. Starobinsky, \Journal{\PLB}{117}{175}{1982}.

\bibitem{Martin:2007bw} J. Martin, \Journal{\LNP}{738}{193}{2008}.

\bibitem{PlanckNG} P. Ade et al., {\tt arXiv:1303.5084}

\bibitem{Planck} P. Ade et al., {\tt arXiv:1303.5076}.

\bibitem{Martin:2013tda} J. Martin, C. Ringeval, and V. Vennin, {\tt arXiv:1303.3787}.

\bibitem{Martin2013} J. Martin, C. Ringeval, R. Trotta and V. Vennin, {\tt arXiv:1312.3529}.

\bibitem{Turner:1983he} M. Turner, \Journal{\PRD}{28}{1243}{1983}.

\bibitem{Kofman:1997yn} L. Kofman, A. Linde and A. Starobinsky, \Journal{\PRD}{56}{3258}{1997}.

\bibitem{Podolsky:2005bw} D. Podolsky, G. Felder, L. Kofman and M. Peloso \Journal{\PRD}{73}{023501}{2006}.

\bibitem{Martin:2010kz} J. Martin and C. Ringeval, \Journal{\PRD}{82}{023511}{2010}.

\bibitem{Martin:2012pea} J. Martin, V. Vennin and P. Peter, \Journal{\PRD}{86}{103524}{2012}.

\bibitem{Schwarz:2004tz} D. Schwarz and C. Terrero-Escalante, \Journal{\JCAP}{0408}{003}{2004}.

\bibitem{Martin:2013uma} J. Martin, C. Ringeval and V. Vennin, \Journal{\JCAP}{1306}{021}{2013}.

\bibitem{Kunz:2006mc} M. Kunz, R. Trotta and D. Parkinson, \Journal{\PRD}{74}{023503}{2006}.

\bibitem{Trotta:2008qt} R. Trotta, \Journal{\Contemp}{49}{71}{2008}.

\bibitem{Martin:2010hh} J. Martin, C. Ringeval and R. Trotta, \Journal{\PRD}{83}{063524}{2011}.

\bibitem{Feroz:2011bj} F. Feroz, K. Cranmer, M. Hobson, R. de Austri
  and R. Trotta, \Journal{\JHEP}{1106}{042}{2011}.

\bibitem{Ringeval2013} C. Ringeval, {\tt arXiv:1312.2347}.

\end{thebibliography}
\end{document}